\documentclass[11pt]{amsart}

\usepackage{amsmath,amssymb,amsfonts,fullpage,amsbsy,graphicx}
\usepackage{equation,color}

\newtheorem{theo}{Theorem}
\newtheorem{lemma}[theo]{Lemma}

\newtheorem{conj}[theo]{Conjecture}

\newcommand{\R}{{\mathbb R}}

\newenvironment{Proof}{\removelastskip \vskip12pt plus 1pt \noindent
{\em Proof.\/}\rm }{\hfill$\Box$ \vskip12pt plus 1pt} 
\newcommand{\norm}[1]{\| #1 \|}

\newcommand{\blot}[1]{}

\title{A General Model of Structured Cell Kinetics} 

\author{Michael Grinfeld}
\address{Department of Mathematics and
    Statistics, University of Strathclyde, 26 Richmond Street, Glasgow
    G1 1XH, UK}
 \email{m.grinfeld@strath.ac.uk}

 \author{Nigel Mottram}
 \address{School of Mathematics and Statistics,
   University of Glasgow, University Place, Glasgow G12 8QQ, UK}
\email{Nigel.Mottram@glasgow.ac.uk}

  \author{Jozsef Farkas}
 \address{Division of Computing Science and
   Mathematics, University of Stirling, Stirling FK9 4LA, UK}
 \email{jozsef.farkas@stir.ac.uk}

\begin{document}

\begin{abstract}
  \noindent We present a modelling framework for the dynamics of cells
  structured by the concentration of a micromolecule they contain. We
  derive general equations for the evolution of the cell population
  and of the extra-cellular concentration of the molecule and apply
  this approach to models of silicosis, quorum sensing in
  Gram-negative bacteria and magnetic ion exchange.\\

\end {abstract}

\maketitle

\section{Introduction}

Many physical systems involve the interaction of micro-scale objects
and macro-scale objects within a region. For instance, in biology, the
micro-scale objects could be molecules of a particular chemical with
the macro-scale objects could be cells, and the region could be a
Petri dish or an organ. This region, a domain $\Omega$ of volume $W$,
may be of fixed size or change with time, but we assume that the
micro-scale objects, of a species $X$ (which from now we call
molecules for brevity), have no volume while the macro-scale objects
(which from now on we call cells) each have volume $V_0$. The $X$
molecules may be present inside or outside the cells, with the
concentration of $X$ varying between the cells. We also assume that
the molecules of $X$ can participate in any subset of the following
processes: they can be injected into or be removed from the domain,
they can enter and exit cells and they can be produced, processed and
destroyed by cells. Suppose also that the fate of a cell is dependent
on the amount of $X$ that it contains. The goal of the present paper
is to introduce, using ideas of Metz and Diekmann \cite{MD} and of
Brown \cite{Brown}, a modelling framework for such situations.

Below we describe in detail three specific examples of systems
where our modelling approach is appropriate, the dynamics of
silicosis, the biological background for which can be found in
\cite{Tran}, quorum sensing in Gram-positive bacteria following Brown
\cite{Brown} and magnetic ion-exchange resin water treatment
  \cite{Boodoo}. However, our framework is suitable for many other
situations, some of which are briefly considered in
Section~\ref{Conc}.  We are confident there are many other examples,
in both biological and non-biological systems, where the proposed
philosophy may be useful.

The structure of the paper is as follows. In Section~\ref{Sil} we
introduce the biological background of silicosis and argue that the
mathematical model of \cite{Tran}, which is couched in terms of
coagulation-fragmentation equations, does not give a correct
description of the dynamics; this is our original motivation for
developing the present approach. In Section~\ref{Mod} we show how to
derive the required equations in the general setting. In
Section~\ref{silm} we complete the specification of the silicosis
model. In Section~\ref{QS} we consider quorum sensing in
  Gram-positive bacteria. This is a useful setting for testing
  numerical approaches to the type of models we are interested
  in. Surprisingly, our model also allows a wealth of stationary
  solutions and an intriguing reinterpretation of the whole concept of
  bistability. In Section~\ref{MIEX} we discuss a series of models for
  magnetic ion-exchange resin-based water treatment (MIEX). Finally,
in Section~\ref{Conc} we suggest other areas of application of our
framework, discuss general mathematical issues and draw
conclusions.

Note that the present paper is purely methodological and that all
results on existence, uniqueness, or asymptotic behaviour of solutions
of the type of equations we derive here, are left to future work.

\section{Silicosis: the Coagulation-Fragmentation
  Approach} \label{Sil}

Let us summarise the 1995 silicosis model of Tran {\em et
  al.}~\cite{Tran}, which should be consulted for references. The
biological background is as follows: quartz particles are ingested and
arrive in the lung. There they may be picked up by macrophages, with
the intent of being removed together with their quartz load via the
muco-ciliary escalator. However, if a macrophage accumulates too large
a quartz load, it becomes immobile and eventually dies by apoptosis in
the lung, releasing its quartz load.

The variables in the model of Tran {\em et al.}~are: free quartz dust in
the lungs in concentration $x(t)$ and concentrations of macrophages
$M_k(t)$ containing $k$ {\em particles} of quartz. They write down an
equation for the evolution of free quartz particle concentration and
for $M_k$, equations of the form
\begin{equation}
  \frac{dM_k}{dt} = \alpha_{k-1}x M_{k-1}- \alpha_k x M_k + \cdots ,\label{tran}
\end{equation}
where $\alpha_{k-1}$ and $\alpha_{k}$ are the kinetic constants for
the process of macrophages with $k-1$ and $k$ particles of quartz,
respectively, ingesting one additional quartz particle.

In eq.~(\ref{tran}), the $\cdots$ stand for the two different
``death'' processes: the disappearance of cells together with their
quartz load via the the muco-ciliary elevator; or cell apoptosis
accompanied by the release of the quartz load into the lungs. To make
this model fully specified, it must be complemented by an equation for
the production of na\"{i}ve macrophages, $M_0(t)$; this rate in
general depends on the quartz load in the lungs. See
Section~\ref{silm} below for a reasonable form of such an equation.

This model was later considered in \cite{CDG}; mathematically it is
interesting and falls in the framework of coagulation-fragmentation
equations (see \cite{BLL} for an up-to-date exposition of this area of
infinite-dimensional dynamical systems). The global existence of its
solutions has been proven in later work by da Costa and coauthors
\cite{CPS}, and many mathematical questions connected with the model
of Tran {\em et al.}~are still open.

To understand our objections to the modelling of silicosis as a
coagulation--fragmentation system, consider a typical
coagulation--fragmentation reaction scheme,
\[
  c_k+ c_1 \rightleftharpoons c_{k+1}.
\]
Here $c_k$, $c_1$, $c_{k+1}$ are {\em concentrations} of $k$-mers,
monomers, and $(k+1)$-mers of some chemical species, respectively. The
equation for the evolution of $c_{k+1}$ corresponding to this reaction
scheme is
\[
  \frac{dc_{k+1}}{dt} = \alpha_{k}c_k c_1 - \beta_{k+1} c_{k+1},
\]
where $\alpha_{k}$ is the kinetic constant for the coagulation
reaction between $k$-mers and monomers and $\beta_{k+1}$ is the
kinetic constant of the fragmentation of a molecule of $(k+1)$-mer
into a monomer and a $k$-mer. This equation is simply mass-action
kinetics, and $\alpha_k$ and $\beta_{k+1}$ are assumed to be functions
of $k$ only.

If we now compare this coagulation--fragmentation process with
eq.~(\ref{tran}), we would suppose that the underlying reaction scheme
is
\[
    M_k + x \rightleftharpoons M_{k+1}.
\]  
This situation is subtly different in that the reaction here
is between the molecules of quartz outside the cells and the {\em
  content} of the cells. Though $M_k$, $x$, $M_{k+1}$ have units of
concentration, the reaction encoded in this scheme in general involves
the {\em concentration} of quartz, and not the number of quartz
particles, inside the cells. As an example, consider the case of
exchange of quartz between the outside and the inside of a cell driven
by passive (Fickian) diffusion. Then the rate of exchange of molecules
is proportional to $(x-q_I)$, where $q_I$ is the internal
concentration of quartz.

However, the $k$ in $M_k$ denotes the {\em number} of quartz molecules
in a cell, not their concentration. To define internal concentration
of quartz, we cannot assume that a cell is a point object, and have to
endow it with a volume, say $V_0$. But if cells now have finite
volumes, it becomes clear that the concentration of free quartz is
also a function of the number of cells which is not the case in the
coagulation--fragmentation setup. This is precisely the kind of
confusion of units that is avoided in the type of model proposed
below.

\section{Derivation of General Equations} \label{Mod}

\begin{figure}[htp]
\centerline{\includegraphics[width=0.5\textwidth]{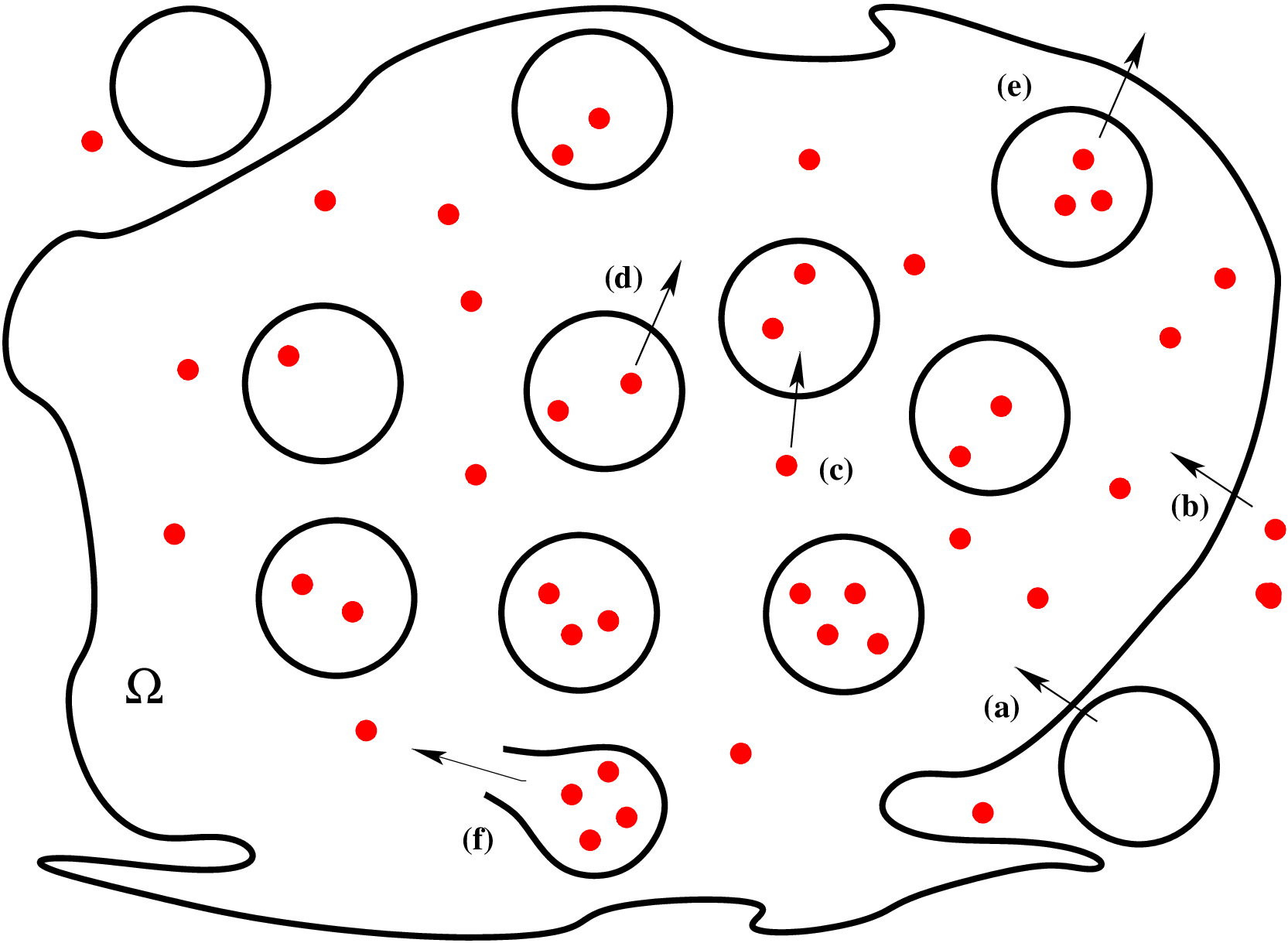}}
\caption{The situation being modelled: the domain $\Omega$ contains
  species $X$ (red dots) and cells (black circles). The possible
  processes involved are: (a) na\"{i}ve cells enter the domain; (b)
  species $X$ enters the domain; species $X$ (c) enters or (d) exits
  cells; (e) $X$-laden cells exit the domain and (f) cells release
  molecules of $X$ in $\Omega$; also present, but not indicated in
  this figure, are synthesis and degradation of $X$.}
\label{fig:1}
\end{figure} 

We are modelling the situation sketched in Figure~\ref{fig:1} and
described as follows. The system consists of a domain $\Omega$ of
volume $W$ that contains molecules of $X$ (denoted by red dots in
Fig.~\ref{fig:1}) with extracellular concentration $x$, and a
population of cells (the black circles in Fig.~\ref{fig:1}). We assume
that the volume of each cell is $V_0$ and treat the molecules as
having no volume.

The cells differ in their $X$ content, and we define the
time-dependent density of the cells with internal concentration of $X$
being $y$ to be $M(y,t)$. That is, the number of cells with internal
concentrations of $X$ between $y_1$ and $y_2$, with $y_1 < y_2$, is
  \[
    {\displaystyle \int_{y_1}^{y_2} M (y, t) \, {\rm d}y}.
   \] 
Then 
\[
  V(t)= V_0 \int_0^\infty M (y, t) \,{\rm d}y
\]
is the total volume occupied by the cells and we assume that for all
time $t$, $V (t) < W$. By this we mean that if $V(0) < W$, the
dynamics of the system ensures that $V(t) < W$.

Furthermore, inside each cell we have a process that involves the
substance $X$, which can be described by an ordinary differential
equation such as
 \begin{equation}\label{ieq}
   \frac{dy}{dt} = f (x, y) = J(x,y) + g(y),
 \end{equation}
 where we have separated the transport term $J(x,y)$, that may depend
 on both the internal and external concentration of $X$, and the
 intracellular synthesis and degradation term $g(y)$. This type of
 equation is called an $i$-equation (for individual) by Metz and
 Diekmann \cite{MD}.

 Metz and Diekmann \cite{MD} also show (in many different ways) how to
 derive the equation governing the evolution of $M(y,t)$, what they
 call the $p$-equation (for population), 
 \begin{equation}\label{peq}
   M_t (y, t) + (f (x, y)M (y, t))_y = P (M (y, t), y) + Q(M(y,t),y),
 \end{equation} 
 where in the right-hand side the terms $P$ and $Q$ encode all the
 population level processes (such as birth and death). Specifically,
 we denote by $Q(M (y, t), y)$, the population level processes that
 feed back into the $x$ dynamics (e.g.~when cells die in $\Omega$
 releasing their contents). The model is then closed by specifying the
 $x$ dynamics, adding suitable initial
 conditions $x(0)$ and $M(y,0)$, and a boundary condition for
 na\"{i}ve cells entering $\Omega$, usually of the form
\[
  f (x(t),0) M (0, t) = s(\cdot),
\]
where the function $s$ can depend on a variety of variables. Thus the
unknown dependent variables are $x(t)$ and $M(y,t)$.

Deriving the equation governing the evolution of $x(t)$ is
algorithmic, by keeping account of the total number of molecules of
$X$ in the extracellular space. In general, the result is an
integro-differential equation. The derivation of the equation for
$x(t)$ crucially uses the thinking of Brown \cite{Brown}, which we now
discuss.

Brown \cite{Brown} considers the system described above but with the
simplification that every cell contains the same number of molecules,
so that the concentration of $X$ in all cells is $y$, a constant.
If $K(t)$ is the number of cells at time $t$, then the total volume
occupied by cells is $V(t)=K(t)V_0$, the cell-free volume is
$W-V(t)$ and the total number of molecules outside of cells is
$N_E=(W-V(t))x(t)$.

We will work in two stages. First we write equations for the transport
of $X$ molecules between the interior and exterior of the cells,
assuming that the flux is given by $J(x,y)$ and that it is positive
when molecules enter the cells. We have
\begin{equation}\label{eqne}
      \frac{dN_E}{dt} = -VJ(x,y).
\end{equation}
An obvious example of a flux would be $J(x,y)=D(x-y)$, with $D$
a diffusion constant, but in examples of interest, mechanisms
involving facilitated transport or phagocytosis should be  considered.  

The proportionality to $V$ in eq.~(\ref{eqne}) comes from the
consideration that the rate of change should be proportional to the
available surface area of the cells, which, given we assume a fixed
individual cell volume and surface area, is proportional to the number
of cells and thus proportional to the total volume of cells.

Now let us rewrite these equations in terms of concentrations
only. Since $N_E=x(W-V)$, we have 
\[
    \frac{dx}{dt} = -\frac{V}{W-V} J(x,y) + \frac{x}{W-V}
    \frac{dV}{dt}.
\]
Now we add to this molecular transport equation terms involving
synthesis and degradation terms which we collect in one term,
$H(x,y)$; we have 
\begin{equation}\label{eqbr}
         \frac{dx}{dt} = -\frac{V}{W-V} J(x,y) + \frac{x}{W-V}
         \frac{dV}{dt} + H(x,y).
\end{equation}
Here $H(x,y)$ incorporates all extracellular production and
degradation of $X$ and all the population level processes that feed
back into the extracellular concentration $x(t)$.

We now consider the more general case in which the cells may have
different internal concentrations of $X$. This necessitates a number
of changes. First of all, the total number of molecules being released
per unit time by cellular processes, i.e~$Q$ in eq.~(\ref{peq}), is
${\displaystyle V_0 \int_0^\infty y Q(M(y,t),y) \, dy}$ and hence we
can write 
\[
  H(x,y) = h(x) + \frac{V_0}{W-V} \int_0^\infty y Q(M(y,t),y) \, dy,
\]
where $h(x)$ is the rate of extracellular production and
degradation of $X$, which depends on the particular modelling context. 

Secondly, the term $V(t)J(x,y)=V_0 K(t) J(x,y)$ is replaced by the
integral ${\displaystyle V_0 \int_0^\infty M(y,t) J(x,y) \, dy}$. With
these changes, the equation for the evolution of $x(t)$ becomes   
 \begin{equation}\label{newx}
    \frac{dx}{dt} = h(x) + \frac{1}{W-V} \left[
      -V_0\int_0^\infty J(x,y)M(y,t) \, dy\right. + \left. V_0\int_0^\infty
     y\,Q(M(y,t),y)\, dy + x \frac{dV}{dt} \right].
 \end{equation} 

 Therefore our modelling framework consists of the equation governing
 the cell population, the $p$-equation, eq.~(\ref{peq}); the equation
 governing the concentration of $X$ external to the cells, the
 $x$-equation, eq.~(\ref{newx}); and the data for any particular model
 set through the biologically determined terms $J(x,y)$, $g(x)$,
 $P(M(y),t), y)$, $Q(M(y),t), y)$, $h(x)$ and the boundary condition
 function $s(\cdot)$.

 \section{A New Model for Silicosis}\label{silm}
 
In this section we formulate a model of silicosis based on the
  principles of Section~\ref{Mod}. As in Tran {\em et
  al.}~\cite{Tran}, we assume that quartz is being ingested at a
constant rate. We set $M(y,t)$ to be the density of macrophages having
internal quartz concentration $y$ at time $t$.  As in \cite{Tran}, we
assume that new macrophages are produced at rate $s$ determined by the
quartz load, ${\displaystyle L(t) := \int_0^\infty yM(y,t)\, dy}$,
such that
\begin{equation}\label{eqs}
  s(L(t)) = s_0 + u(L(t)),
\end{equation}
where $s_0$ is a background level of recruitment of na\"{i}ve cells
into the domain when quartz is not present in any cells and $u(\cdot)$
is a bounded function, with $u(0)=0$.

Cells with internal concentration of quartz $y$ are removed by the
muco-ciliary escalator at a rate $p(y)$, where $p(y)$ is a decreasing
function of $y$ since, as their quartz content increases, cells are
increasingly immobile. Cells are also more liable to die by apoptosis
as their quartz content increases, and so the rate of them releasing
their contents inside the lungs, $q(y)$, is an increasing function of
$y$.

As there is no intracellular processing of quartz, so that $g(y)=0$ in
eq.~(\ref{ieq}), we only need to specify the transport
mechanism. Phagocytosis of quartz particles cannot be described by
simple diffusion, so we set
\begin{equation}\label{i2}
  \frac{dy}{dt} = J(x,y), 
\end{equation}
where the function $J$ is non-negative, bounded, increasing in
$x$ and decreasing in $y$, as is also assumed in \cite{Tran}. An
example would be
\[
  J(x,y) = \frac{\gamma x}{x+y+x_{1/2}},
\]
where $\gamma$ is the flux when $x\rightarrow\infty$ and $x_{1/2}$ at
which the value of $x$ when the flux for na\"{i}ve cells (i.e., $y=0$)
is half the maximum value, both positive constants.

So far we have all the information needed to specify the $p$-equation
for $M$, which is therefore
\begin{equation}\label{p2}
M_t(y,t) + (J(x,y) M(y,t))_y = - (p(y)+q(y))M(y,t). 
\end{equation}

Now we need to formulate the equation for $x(t)$. From (\ref{newx}) it
follows that all we need to do is to specify the function $h(x)$, the
rate of change of concentration of quartz particles in the
extracellular region due to introduction from outside the domain. If
we assume that $A$ particles of quartz are ingested per unit time, we
have
\[
h(x) = \frac{A}{W-V},
\]
and hence 
\begin{equation}\label{eqx}
  \frac{dx}{dt} = \frac{1}{W-V} \left[ A - V_0\int_0^\infty
    J(x,y)M(y,t)
    \,dy + 
    V_0\int_0^\infty yq(y)M(y,t)\, dy + x\frac{dV}{dt} \right].
\end{equation}

We can derive an expression for $dV/dt$ in terms of the variables
$M(y,t)$ to substitute in  (\ref{eqx}). Since $dV/dt$ is given by
\[
  \frac{dV}{dt} = V_0\int_0^\infty M_t(y,t) \, dy,
\]  
integrating the $p$-equation by parts and assuming that 
that for all times $t>0$ we have that
\[
  \lim_{y \to \infty} J(x,y) M(y,t)=0,
\]
we obtain
\begin{equation}\label{dvdt}
  \frac{dV}{dt} = V_0\left(s(L(t)) -\int_0^\infty (p(y)+r(y)) M(y,t)\,
    dy\right).
\end{equation}

In addition, we must specify suitable initial conditions $x(0)$ and
$M(y,0)$ as well as the boundary condition at $y=0$, which, using
eq.~(\ref{eqs}), is
\begin{equation}\label{bc}
  J(x,0)M(0,t)= s_0 + u(L(t)).
\end{equation}

We note that the resulting system is a linear transport equation (for
the population variable $M$) coupled to an nonlinear
integro-differential differential equation for the extracellular
quartz concentration $x$.

Having assumed that the load
\[
    L(t)= \int_0^\infty y M(y,t) \, dy
\]
is finite for all time, the correct setting for the theory is a space
of positive Radon measures with a finite first moment. Particular
choices of the functions $p(y)$ and $r(y)$ and the input function
$s(\cdot)$ must ensure that if $V(0)<W$, then $V(t)<W$. For example,
this can be shown to be the case if we assume that $p(y)+r(y)$ is
bounded below and that $s(\cdot)$ is bounded above by constants.

\section{Quorum sensing in Gram-negative bacteria}\label{QS}

\subsection{Derivation of equations}

A number of models for quorum-sensing in Gram-positive and
  Gram-negative bacteria that assume that the internal concentration
  of the signal molecule $X$, i.e., $y$, is the same in all cells, see
  \cite{Brown}; we will review the model for Gram-negative bacreria
  below.  The case of Gram-negative bacteria is an interesting test
  case of our approach, as, like the case of magnetic ion-exchange
  resin-based water treatment discussed in section \ref{MIEX}, it
  involves a significant simplification: we can assume that the number
  of cells is constant, so $V \equiv \hbox{const}$ is now a parameter
  and $dV/dt=0$.  Below, as is done by Brown, we will find
  convenient to use the parameter $\phi=V/W$. Fickian diffusion of $X$
  between the cells and the extracellular region is assumed, so that
  the $i$-equation is
\begin{equation} \label{yeqqs}
  \frac{dy}{dt}=F(x,y) := D(x-y)+g_s(y)+g_d(y),
\end{equation}
where $D$ is a diffusion constant, and following \cite{Brown},
we include terms modelling intracellular synthesis $g_s(y)$, which we
take to be a constant term $a_0$ plus a Hill-type term,and a
  linear intracellular degradation term $g_d(y)$. If we choose the
  Hill exponent to be $2$, we have
\begin{equation}\label{fxy}
F(x,y)= D(x-y)+ a_0 + \frac{a_1 y^2}{K^2+y^2} - m_Iy,
\end{equation}
for some values of the constants $a_1$ and $K$. The $p$-equation is
\begin{equation}\label{meqqs}
  M_t(y,t) + (F(x,y)M(y,t))_y =0.
\end{equation}
The boundary condition is $M(0,t)=0$ and the $x$ equation becomes 
\begin{equation}\label{xeqqs}
  \frac{dx}{dt} = \frac{V_0 D}{W-V}\int_0^\infty (y-x)M(y,t)\, dy - 
m_E x, 
\end{equation}
where $m_E$ is the extracellular degradation rate. 

\subsection{Analysis of equilibria}

The set of equilibria of the model given by
  (\ref{yeqqs})--\ref{xeqqs}) is quite rich. Analysis of this set
  changes our view of bistability, and so is definitely worth the
  effort. There are many cases to consider, and instead of an
  exhaustive analysis left to the reader, we present a classification
  of possible cases and analyse in detail a representative one.

First, assuming that all the cells have the same concentration
  $y$ of the signal molecule, we have the model of Brown as in
  \cite{Brown}, Equations (3) and (in non-dimensional form) (4). The
  relevant results of numerically solving the steady state equations,
  are in \cite[Fig. 1(B)]{Brown}. The blue curve there corresponding
  to Hill exponent $2$, which shows the dependence of the internal
  concentration of the signal on $\phi$ clearly shows bistability as
  it is commonly understood: for small enough $\phi$ there is a unique
  equilibrium with low concentration of $y$; call this branch of
  equilibria $y_L\phi)$. Then there is a saddle-node bifurcation at
  $\phi_1$ in which two more branches of equilibria are born, call
  them $y_I(\phi)$ and $y_H(\phi)$ (for {\em I}ntermediate and {\em
    H}igh, respectively); finally, $y_I(\phi)$ and $y_L(\phi)$ collide
  in yet another saddle-node bifurcation at $\phi_2> \phi_1$ and only
  the $y_H(\phi)$ branch remains. In the (bistability) parameter
  regime $\phi_1 < \phi < \phi_2$, the system of 2 ODEs
  \cite[Eq. (4)]{Brown} has three equilibria, which in our notation
  can be written as $(x_L(\phi), \, y_L(\phi))$,
  $(x_I(\phi), \, y_I(\phi))$, $(x_H(\phi)\, \, y_H(\phi))$, the first
  and the third of which are locally asymptotically stable, the
  second being a saddle point. Therefore, depending on the initial
  condition in the bistability regime, with probability 1 the solution
  of the system of ODES will converge either to
  $(x_L(\phi), \, y_L(\phi))$ or to $(x_H(\phi), \, y_H(\phi))$.

The situation is much more complex in the case of model
  (\ref{yeqqs})--\ref{xeqqs} and requires a reassessment of what
  bistability means.

 \subsubsection{Notation and preliminaries}

  The analysis is absolutely elementary, but it requires careful
  notation. Consider the equation $F(x,y)=0$. As it is linear in $x$,
  from this equation, we can express $x$ as a function of $y$; thus we
  obtain one relation between $x$ and $y$ that must be satisfied at
  equilibrium. We record this as
\begin{equation}\label{stat1}
  x = g(y) := y - \frac{1}{D} \left[ a_0 + \frac{a_1 y^2}{K^2+y^2} -
    m_Iy \right].
\end{equation}  
Note that $g(y) \sim (1+m_I) y> y$ for large enough $y$. This means
that for any $0<A<1$ the straight line $x=Ay$ will intersect the graph
of $g$. Clearly, we can choose parameters $D, a_0,a_1,K, m_I$ such
that there is an interval of values of $x$ which we denote by
$J:=[x_{min}, \, x_{max}]$ such that if $x_0 \in J$, the equation
$x_0=g(y)$ has exactly 3 positive solutions (and never more). For such
a choice, see \cite[Table 1]{Brown}. We denote these solutions by
$y_l(x_0) <  y_m(x_0) < y_r(x_0)$, for obvious reasons. 

Clearly, we also have $y_m(x_{min}) = y_r(x_{min})$ and
$y_l(x_{max}) = y_m(x_{max})$.

We distinguish two fundamental configurations:

I: ${\displaystyle \frac{x_{min}}{y_l(x_{min})} >
  \frac{x_{max}}{y_r(x_{max})}}$, and 

II: ${\displaystyle \frac{x_{min}}{y_l(x_{min})} <
  \frac{x_{max}}{y_r(x_{max})}}$.
 
\begin{figure}[htp]
\centerline{  \includegraphics[width=0.8\textwidth]{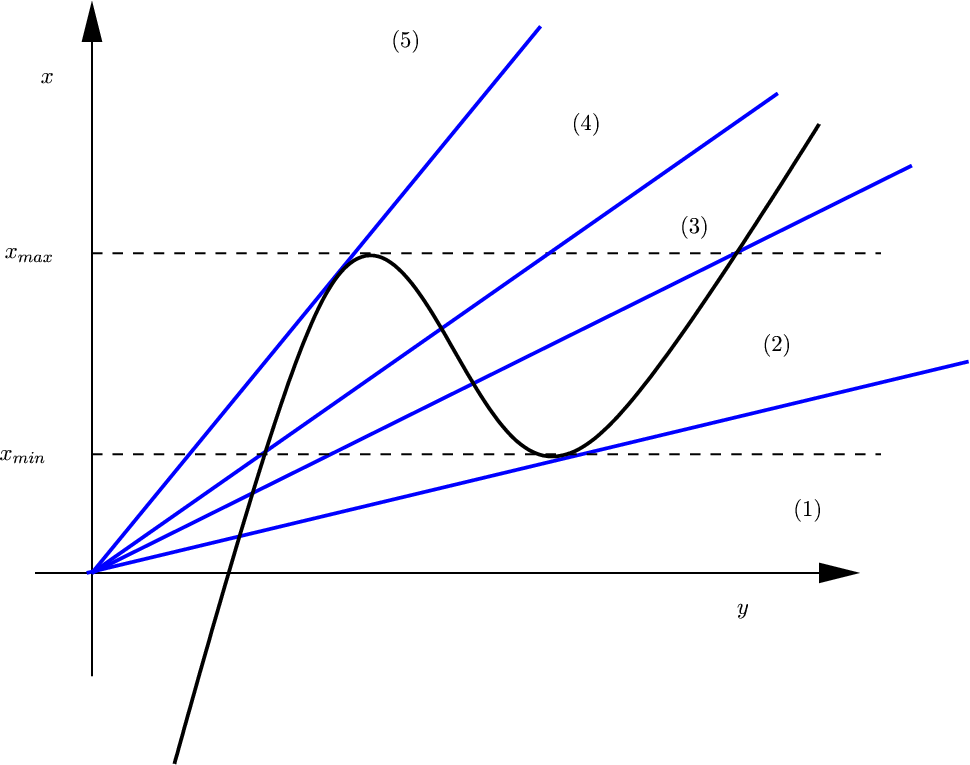} }
\caption{The 5 cases of configuration I.}\label{fig:cases}
\end{figure}

Below we only deal with a particular case of configuration I; all the
other cases are left to the reader. In fact, in configuration I there
are five cases to consider, depending on the intersections of the
lines $x=Ay$ with $x=g(y)$ as we vary $A$. These are indicated in
Figure~\ref{fig:cases}.

Note that in regions (1) and (5) there is no bistability; we
concentrate on region (3), the analysis in regions (2) and (4) is
similar, though the ``bifurcation diagrams'' in a sense to be defined
below are different.

\begin{figure}[htp]
\centerline{  \includegraphics[width=0.8\textwidth]{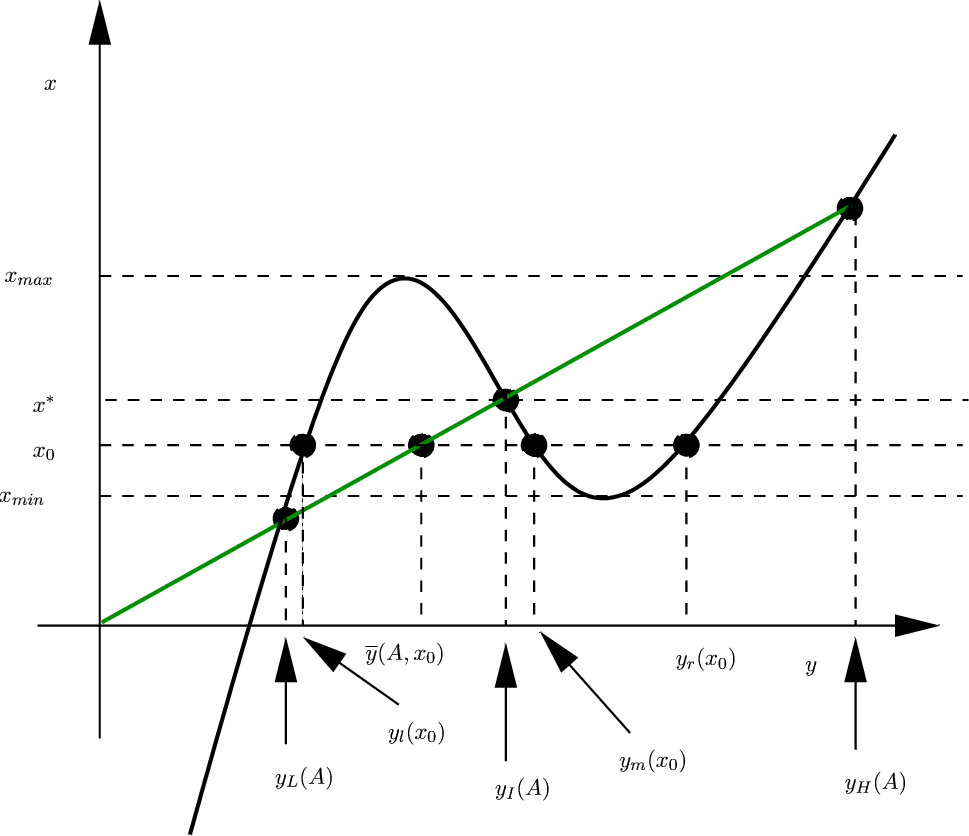} }
\caption{Notation used in the construction.}\label{fig:solns}
\end{figure}

To start constructing stationary solutions of
  (\ref{yeqqs})--(\ref{xeqqs}), pick a slope $A$ in region (3) of
  configuration I. Call the $y$-coordinates of intersections of the
  line $x=Ay$ and $x=g(y)$, $y_L(A) < y_I(A) < y_H(A)$. Let
  $x^*= Ay_I(A)$. For any $x_0 \in J$, as before, denote the
  $y$-coordinates of the intersections of the line $x=x_0$ with the
  graph of $g$ by $y_l(x_0)< y_m(x_0) < y_r(x_0)$. Finally, denote the
  $y$-coordinate of the intersection of the line $x=x_0$ the line
  $y=Ax$ by $\overline{y}(A,x_0)$ to emphasise its dependence both on
  $A$ and $x_0$. See Figure~\ref{fig:solns}.

The solutions we will be constructing for $M(y)$ are sums of
  Dirac deltas at some locations $y_1, y_2,$ etc. In fact we will show
  that there can be only 3 terms in such a sum. If there is only one
  term in the sum, then we call such a solution a $1$-$\delta$ solution,
  if there are $2$, a $2$-$\delta$ solution, and if $3$, a $3$-$\delta$
  solution.

Note that to satisfy the volume constraint, we must have that
such an $n$-$\delta$ solution has the form
\begin{equation}\label{My}
  M(y) = \frac{V}{V_0}\sum_{k=1}^n \alpha_k \delta_{y_k}(y), \quad
  \alpha_k \geq 0, \quad \sum_{k=1}^n\alpha_k=1.
\end{equation}  

Before we start, we prove a simple lemma. 

\begin{lemma}\label{del}
  If for some $x=x_0$, $M(y)$ has the representation (\ref{My}), then
  $x_0=g(y_k)=0$ and hence $n \leq 3$.
\end{lemma}

\begin{Proof} Since in that as in the sense of distributions
  $(F(x_0,y)M(y))_y=0$, multiplying by a test function $u$ on $\R_+$ and
  integrating over $y$ we have that
  \[
    \sum_{k=1}^n F(x_0,y_k)\alpha_k u(y_k)=0,
  \]
 and this can only be true for all test functions $u$ if
 $F(x_0,y_k)=0$. The second claim follows as $x_0=g(y)$ has at most
 $3$ solutions by our assumptions on $F(x,y)$.
\end{Proof}

(a) $1$-$\delta$ solutions: Using Lemma~\ref{del}, these are
  simply obtained by computing the $y$-coordinates of the intersection
  of the graph of $g$ defined in (\ref{stat1}) with the straight line
  $x=Ay$, where from (\ref{xeqqs}) we have that
  \begin{equation}\label{Aeq}
    A= \frac{\phi D}{m_E(1-\phi)+ \phi D}.
  \end{equation}  
  Thus $0\leq A \leq 1$ and $A$ is a monotone increasing function of
  $\phi$. Note that in region (3) of configuration I, we have that
  \[
    \frac{x_{max}}{y_r(x_{max})} \leq A \leq
  \frac{x_{min}}{y_l(x_{min})}.
\]
  
To summarise: for $A$ in region (3) of configuration I there are
three $1$-$\delta$ solutions with support in $y_L(A)$, $y_I(A)$ and
$y_H(A)$; the corresponding values of $x$ are $Ay_L(A)$, $Ay_I(A)$ and
$Ay_H(A)$, respectively.

Thus the set of $1$-$\delta$ equilibrium solutions is exactly the same
as the set of equilibria in the ODE model of Brown \cite{Brown}. We
also expect $(V/V_0\delta_{y_L(A)}, \, A y_L(A))$,
$(V/V_0\delta_{y_H(A)}, \, A y_H(A))$ to be stable in some definable
sense, while $(V/V_0\delta_{y_I(A)}, \, A y_I(A))$ should not be
stable, as in the ODE case. Of course if we are in regions (1) or (5)
of Figure~\ref{fig:cases} there exists a unique $1$-$\delta$ solution
and this exhausts the whole set of equilibria.

(b) $2$-$\delta$ solutions: Here we will show that for {\em
    each} value of $x_0 \in J$ there are two $2$-$\delta$ solutions
  and globally in region (3) of configuration I they form a
  continuum. Note that the construction is made for a fixed $A$,
  i.e. by (\ref{Aeq}), for a fixed value of $\phi$, and we do not
  attempt to make a global bifurcation diagram using $\phi$ (or $A$)
  as a bifurcation parameter.

  Pick any value of $x_0$; in Figure~\ref{fig:solns} we chose
  $x_0 < x^*$; the other case is similar. Then the only allowable
  supports of the Dirac deltas are at $y_l(x_0)$, $y_m(x_0)$ and
  $y_r(x_0)$. At the same time, we have $A\overline{y}(A,
  x_0)=x_0$. As for $x_0<x^*$, and so
  $\overline{y}(A, x_0) \in (y_l(x_0), y_m(x_0))$, we can satisfy all the
  conditions for an equilibrium by choosing $r_1 \in (0,1)$ so that
  $r_1y_l(x_0)+(1-r_1)y_m(x_0)=\overline{y}(A,x_0)$ or $r_2 \in (0,1)$ so
  that $r_2y_l(x_0)+(1-r_2)y_r(x_0)=\overline{y}(A,x_0)$, which is clearly
  possible by the geometry of Figure~\ref{fig:solns}.  But this means
  that for any fixed value of $x_{min}< x_0< x^*$ we have constructed
  two $2$-$\delta$ solutions,
  \[
    M_1(y) = \frac{V}{V_0} \left[r_1
      \delta_{y_l(x_0)}(y)+(1-r_1)\delta_{y_m(x_0)}(y) \right] \hbox{ and
    } M_2(y) = \frac{V}{V_0} \left[r_2 \delta_{y_l(x_0)}(y)
      +(1-r_2)\delta_{y_r(x_0)}(y) \right],
\]
and there can be no others. Note that the dependence of these
solutions on $A$ is via $\overline{y}(A,x_0)$ and hence the constants
$r_1$ and $r_2$.  Clearly, as $x_0 \rightarrow x_{min}$ from above,
the two solutions approach each other and disappear in a
``saddle-node'' bifurcation at $x_0=x_{min}$.

   For $x_0>x^*$, the situation is similar, and we still have exactly
   two  $2$-$\delta$ solutions,
  \[
    M_1(y) = \frac{V}{V_0} \left[r_1 \delta_{y_l(x_0)}(y)
      +(1-r_1)\delta_{y_r(x_0)}(y) \right] \hbox{ and } M_2(y) =
    \frac{V}{V_0} \left[r_2
      \delta_{y_m(x_0)}(A)+(1-r_2)\delta_{y_r(x_0)}(y)\right],
\] 
and again these two solutions disappear in a ``saddle-node''
bifurcation as $x_0 \rightarrow x_{max}$ from below.

We conjecture that such mixture solutions are ``stable'' if their
support does not include $y_m(x_0)$ and unstable if it does.

Finally, at $x_0=x^*$, one of these two equilibria degenerates into a
$1$-$\delta$ solution supported at $y_m(x^*)=y_I(A)$. 

Note that for a fixed $A$ we have constructed a {\em closed} curve of
$2$-$\delta$ equilibria in the space $\R \times {\mathcal M}_{V/V_0}^2$,
where ${\mathcal M}_{V/V_0}$ is the space of positive measures of mass
$V/V_0$. Visualising it in $\R^3$ is an interesting question.

(c) $3$-$\delta$ solutions. Clearly, for each $x_0 \in J$ we can
  find $r,s \in [0,1]$ such that
  \[
    \overline{y}(A,x_0) = r y_l(x_0)+s y_m(x_0)+ (1-r-s)y_r(x_0),
  \]
  which will correspond to a $3$-$\delta$ solution. The geometry of
  the resulting surface in the $(x_0, r,s)$ space is an intriguing
  question. We expect that all these $3$-$\delta$ solutions will be
  unstable.

  \subsubsection{A remark and a conjecture}. We see that in the system
  (\ref{yeqqs})--(\ref{xeqqs}) the meaning of bistability differs from
  that it usually has in lumped systems. Here it means the coexistence
  of an uncountable number of invariant sets with nonempty domains of
  attraction. These collapse to a unique set as $A$ ($\phi$) increases
  or decreases sufficiently. It is true that individually each of the
  $2$-$\delta$ solutions has a vanishing domain of attraction, but the
  whole set of these solutions collectively has a domain of attraction
  comparable to that of the $1$-$\delta$ solutions supported at
  $y_L(A)$ and $y_H(A)$.

  We have the following conjecture:

  \begin{conj}
    Suppose the initial conditions $(x(0),M(y,0)$ are such
    $M(y,0) \in A$, where $A$ is the set of positive measures $P(y)$
    of mass $V/V_0$, such that the support of $P(y)$ contains
    $y_m(x(0))$. Then the set of elements in $A$ such that the
    solution of (\ref{yeqqs})--(\ref{xeqqs}) does not converge to a
    $2$-$\delta$ stationary solution is null.
  \end{conj}

\section{Magnetic Ion-Exchange Resin-based Water
  Treatment} \label{MIEX}

This section has been written in collaboration with Geraldine
  Knops.

\subsection{Background and a minimal model}

\begin{figure}[htp]
\centerline{  \includegraphics[width=0.8\textwidth]{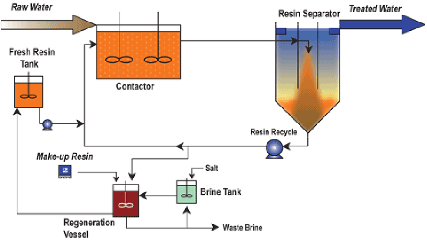} }
	\caption{MIEX setup}\label{fig1} 
\end{figure}

The minimal scheme for MIEX technology is as in
Figure~\ref{fig1}. This figure is taken from \cite{Boodoo}; other
references are in the work of Boyer \cite{Boyer1,Boyer2}.

The model is formulated under the following assumptions:

\begin{itemize}
  \item Regeneration occurs all the time (and not as it really does,
    on breakout) and it is perfect;
  \item There is no loss of resin ever;
  \item All resin beads have the same size.  
\end{itemize}

Set $x$ to be the DOM concentration in the contactor tank and $M(y,t)$
to be the number density of resin beads there. Then if the volume of a
resin bead is $V_0$, the assumption is that the total volume of resin,
\[
  V(t) = V_0 \int_0^\infty M(y,t)\, dy
\]
is constant. Which is good news as it makes equations much simpler,
but what I do not like is that I have to use it to derive the boundary
condition. We will assume that the absorption of DOM particles onto a
bead has kinetics given by some $f(x,y)\geq 0$, where $f(0,y)=0$, and
$f(x,y) \rightarrow 0$ for all $x$ as $y \rightarrow
\infty$. Following our machinery, if the inflow of DOM is at rate of
$A$ particles per unit time (can be $A(t)$, and the outflow of species
$X$ is at rate $\alpha X$, the equations are
\begin{equation}\label{meq}
  M_t(y,t) + \left( f(x,y)M(y,t) \right)_y = -\alpha M(y,t).
\end{equation}
and
\begin{equation}\label{xeq}
  x'=-\alpha x + \frac{1}{W-V}\left[ A-V_0\int_0^\infty f(x,y)M(y,t)\,
  dy \right].
\end{equation}

Now take the $M$-equation (\ref{meq}), multiply it by $V_0$  and
integrate it wrt $y$. Then we have
\[
  V'(t)+ V_0\lim_{y \rightarrow \infty} f(x,y)M(y,t) - V_0f(x,0)M(0,t)=
  -\alpha V.
\]
Since by assumption $V'(t)=0$, and we can (can we? I guess the answer
is yes if we are in a $y$-weighted $L^1$ space as we should be) assume
that the limit is 0, we have that
\[
  V_0f(x,0)M(0,t)=\alpha V  \in  \R,
\]
as $V$ is just a parameter; this time-dependent boundary condition can
be written as
\begin{equation}\label{bc1}
  M(0,t)= \frac{\alpha V}{V_0f(x(t),0)},
\end{equation}
and it makes sense that increasing
$x(t)$, the effluent concentration of DOM means we will have fewer
immaculate beads $M(0,t)$.

\subsection{Stationary solutions in the minimal model}

Let us assume we scale variables so that
\[
  f(x,y)=\frac{x}{x+y+1},
\]
Set $\int_0^\infty M(y)\, dy = s$.

Then solving the $M(y)$ equation with the boundary condition
\[
  M(0)= \frac{\alpha s}{f(x,0)},
\]
  with $f$ as above, we have
\[
  M(y) = \frac{\alpha s}{x} (x+y+1) {\displaystyle
    e^{-\frac{\alpha y(y+2x+2)}{2x}}}.
\]

Check: the boundary condition is satisfied and 
\[
  \int_0^\infty M(y)\, dy = s.
\]

Let us introduce some additional notation. Let
\[
  g(\alpha,x) = 1-\hbox{erf}\, \left( \sqrt{\frac{\alpha}{2x}}(x+1)
  \right),
\]
Notice that
\[
	\int_0^\infty y M(y)\, dy = \sqrt{\frac{x\pi}{2\alpha}} \exp \left(
      \frac{\alpha(x+1)^2}{2x} \right) g(\alpha, x) < \infty
\]
for any values of parameters, so that we stay in
$L^1((0, \infty),y) $. $\int_0^\infty y M(y)\, dy$ is an increasing
function of $x$ with and we have
  \[
    \lim_{x \rightarrow \infty} \int_0^\infty y M(y)\, dy
    =\frac{s}{\alpha},
  \]
  which makes sense. 

  Now we need to say something about $x$. So
  first we need $\int_0^\infty f(x,y) M(y)\, dy$. This is computable.
  We have
\[
 \int_0^\infty f(x,y)M(y)\, dy = s\sqrt{\frac{\alpha \pi x}{2}}
  g(\alpha,x)\exp \left( \frac{\alpha(x+1)^2}{2x} \right) := s H(x); 
\]
by inspection $H(x)$ is a monotone increasing function,
\[
  \lim_{x \rightarrow 0} H(x)=0, \quad \quad \lim_{x \rightarrow
    \infty}
  H(x)= 1.
\]
Hence the $x$ coordinate of the stationary solution solves the
equation
\[
  \frac{A}{W-V}-\alpha x = \frac{V}{W-V}H(x),
\]
and from the monotonicity properties of $H(x)$ we see that
we have proved the following lemma:

\begin{lemma}\label{un}
  For every value of the parameters $A>0$, $\alpha>0$, $V>0$ there
  exists a unique stationary solution of (\ref{xeq}) and (\ref{meq}), with
  the boundary condition (\ref{bc1}).
\end{lemma}

Furthermore this stationary solution is an increasing function of $A$
as it should be.

Let us fix all the parameter apart from $V$. The we can denote the
unique equilibrium concentration of $x$ by $x(V)$.  We have the
following lemma.
\begin{lemma}\label{inc}
  The total amount of DOM at equilibrium in the contactor tank,
  $Y(V)= \alpha x(V)(W-V)$, is  decreasing function of $V$.
\end{lemma}

\begin{Proof}
  First of all, $Y(V)=A-VH(x(V))$. Hence
  \[
    \frac{dY}{dV}(V)= -H(x(V))-VH'(x(V)) \frac{dx}{dV}(V).
  \]
  Therefore if it so happens that ${\displaystyle \frac{dx}{dV}(V)
    \geq 0}$, we necessarily have that ${\displaystyle \frac{dY}{dV}(V)
    < 0}$. On the other hand, since by definition
  \[
\frac{dY}{dV}(V) = -\alpha x(V) + \alpha (W-V) \frac{dx}{dV}(V),
  \]
  if ${\displaystyle \frac{dx}{dV}(V) < 0}$, we also have that
  ${\displaystyle \frac{dY}{dV}(V) < 0}$.
\end{Proof}

\subsection{A non-local imperfect regeneration model}

A different, more realistic model no longer assumes that
regeneration is perfect.

For $x$, we still have the same equation, (\ref{xeq}). Now let us
introduce a parameter $\beta>1$, which is efficiency of cleaning of
the beads from the DOM gunk. For now let us assume that the
regeneration is instantaneous. So if a bead has concentration $y$ of
DOM, regeneration instantaneously creates (no losses) a bead with
concentration $y/beta$. This seems to me the easiest. That means that
now the $M$ equation is
\begin{equation}\label{meq1}
  M_t(y,t) + \left( f(x,y)M(y,t) \right)_y =
  -\alpha M(y,t) +\alpha M(\beta y,t).
\end{equation}
This is consistent: perfect regeneration corresponds to
$\beta=\infty$ and for all $y>0$,
\[
  \lim_{\beta \rightarrow \infty} M(\beta y,t) =0.
\]  
Now take the $M$-equation (\ref{meq1}), multiply it by $V_0$  and
integrate it wrt $y$. Then we have
\[
  V'(t)+ V_0\lim_{y \rightarrow \infty} f(x,y)M(y,t) - V_0f(x,0)M(0,t)=
  -\alpha V + V_0 \alpha \int_0^y M(\beta y,t) \, dy =
- \alpha V \left(1-\frac{1}{\beta} \right).   
\]
Since by assumption $V'(t)=0$, and we can  assume
that the limit is 0, we have that
\[
  V_0f(x,0)M(0,t)=\alpha \left(1-\frac{1}{\beta} \right) V  \in  \R,
\]
as our new time-dependent boundary condition. As before, it can be
written as
\begin{equation}\label{bc2}
  M(0,t)= \frac{\alpha V}{V_0f(x(t),0)}\left(1-\frac{1}{\beta} \right).
\end{equation}
So the boundary condition transforms very nicely. However, the
$M$-equation (\ref{meq1}) though still linear is now intractable, and
the analysis of stationary solutions no longer holds. Obviously, one
could do a continuation argument in $\epsilon=1/\beta$ as
$\epsilon=\rightarrow 0$.

\subsection{A two tank model}   

In this subsection we consider a completely new type of model,
  which we will call a two-tank model. This type of models seems to be
  very promising.


The $x$ equation is still (\ref{xeq}). What we do next is to
consider two tanks: the contactor tank and the regeneration
tank. Call the number densities of beads in the contactor tank
$M(y,t)$ as before and the number densities of the ones in the
regeneration tank $N(y,t)$. Let the cleaning $i$-equation be
\begin{equation}\label{clean}
  y' = -c(y),
\end{equation}
For example, let us take $c(y)=\beta y$.

The $M$ equation now is 
\begin{equation}\label{ttmeq}
  M_t(y,t) + \left( f(x,y)M(y,t) \right)_y = -\alpha M(y,t) +\alpha N(y,t),
\end{equation}
and the $N$ equation is
\begin{equation}\label{ttneq}
  N_t(y,t) - \left( c(y)N(y,t) \right)_y = \alpha M(y,t) -\alpha N(y,t),
\end{equation}
We assume thus 
$c(y) \sim y$, so $c(0)=0$ (nothing to clean).  Therefore adding
(\ref{ttmeq}) and (\ref{ttneq}), multiplying by $V_0$ and integrating wrt
$y$ between 0 and infinity, we have (since
\[
  V= V_0\left( \int_0^\infty M(y,t)\, dy + \int_0^\infty N(y,t)\, dy
  \right)
\]
is constant) that
\[
  \begin{eqalign}
  V'(t)&~ + V_0\lim_{y \rightarrow \infty} f(x,y)M(y,t) -
  V_0f(x,0)M(0,t)\\
  &~ - V_0 \lim_{y \rightarrow \infty} c(y)N(y,t) +
  V_0 \lim_{y \rightarrow 0} c(y)N(y,t) = 0.
  \end{eqalign}
\]
Since $V'(t)=0$, and we can assume that 
both limits as $y \rightarrow \infty$ are 0, we have that
\begin{equation}\label{bcm}
f(x,0) M(0,t) -  V_0 \lim_{y \rightarrow 0} c(y)N(y,t) = 0 \hbox{ for
  all } t>0.
\end{equation}
Note that $f(x,0)>0$. 

Now we need to think. Since
\[
  \int_0^\infty N(y,t) \, dy < \infty  \hbox{ and }
  \int_0^\infty y N(y,t) \, dy < \infty, 0,
\]
for our choice of $c(y)$, we must have $N(y,t)$ go slower than
$y^{-1}$ as $y \rightarrow 0$. This means that the limit in (\ref{bcm})
must be zero.  Therefore the boundary condition for $M(y,t)$ is
\begin{equation}\label{bcm1}
  M(0,t)=0.
\end{equation}
and for $N(y,t)$ we have
\begin{equation}\label{bcn}
  \lim_{y \rightarrow 0} c(y)N(y,t) =0.
\end{equation}  
We also have
\[
  \int_0^\infty (M(y,t)+N(y,t)) \, dy = V/V_0.
\]
Let us now deal with the (unique) equilibrium.  By adding the
equations we have\footnote{Actually we need to prove that the lhs here
  is continuous on $[0,\infty)$.}
\[
  f(x,y)M(y)-c(y)N(y)=0,
\]  
which allows us to solve for $N(y)$ in terms of $M(y)$ (thus not
having to divide by $y$!):
\begin{equation}\label{mn}
  M(y)= \frac{c(y)N(y)}{f(x,y)}. 
\end{equation}
So now we can substitute (\ref{mn}) into the equilibrium $N(y)$
equation, solve it generating a constant of integration, find $M(y)$
from (\ref{mn}), and then get the constant from the integral
constraint. But that is what Nigel has already done. By construction,
for every $x$ the solution is unique. Then need to substitute in the
$x$ equation and find its value(s) (probably unique as well).

The equilibrium $N(y)$ equation is
\begin{equation}\label{eqn}
  (-c(y)N(y))_y -\alpha(M(y)-N(y))=0,
\end{equation}
integrating which we immediately have that at equilibrium
\[
  \int_0^\infty M(y)\, dy = \int_0^\infty N(y)\, dy,
\]  
which is to be expected. Solving (\ref{eqn}), we have
\[
  N(y) = C y^{\alpha/\beta-1} \exp \left( - \frac{ \alpha
      y(2x+2+y)}{2x} \right),
\]
which shows $N(y)$ is in the right space and that so is $M(y)$ is as
well by (\ref{mn}). Note that as Nigel said, if
$\alpha>\beta$, $N(0)=0$. Here $C$ is a constant of integration to be
found later. Then $M(y)$ is given by
\[
  M(y) = C \frac{y^{\alpha/\beta}\beta (x+y+1)}{x}
  \exp \left( - \frac{\alpha y(2x+2+y)}{2x} \right).
\]  
If we set
\[
  I(\alpha,\beta,x)= \int_0^\infty y^{\alpha/\beta-1} \exp \left( - \frac{ \alpha
      y(2x+2+y)}{2x} \right) \, dy,
\]  
which can be computed in Maple in general in terms of Gamma and Laguerre
functions\footnote{If $\alpha=\beta$, the integral is in terms of the
  error function.}, we have finally that
\begin{equation}\label{C1}
  C=\frac{V}{2 V_0 I(\alpha,\beta,x)}.
\end{equation}

\subsection{Equilibria in the two-tank model}

From now on we simplify to $\beta=\alpha=1$; everything can be done in
full generality but the formulae are very unpleasant.
Defining
\[
	g(x)= \exp \left( \frac{(x+1)^2}{2x} \right) \hbox{ and }
  h(x) = \hbox{erf} \, \left(\frac{x+1}{\sqrt{2x}}\right),
\]
  we have
\[
    I(1,1,x) = \sqrt{\frac{\pi x}{2}}g(x)(1-h(x)), 
\]
and hence using (\ref{C1}), we have the equation for $x$ ($\alpha=1$),
\begin{equation}\label{xeq1}
  \frac{A}{W-V} - x = V H(x),
\end{equation}
where
\[
  H(x) = \frac{\sqrt{\frac{x\pi}{2}}(1+x)g(x)(h(x)-1)+x}{2 I(1,1,x)}.
\]
This function $H(x)$ is by inspection non-negative and monotone
increasing,
\[
  \lim_{x \rightarrow 0} H(x)=0, \quad \quad \lim_{x \rightarrow
    \infty} H(x)=1/2,
\]
so by (\ref{xeq1}), for every value of $V$ we have a unique solution,
which we can call $x^*$, and hence our system has a unique
equilibrium. The argument proving that $x^*(W-V)$ is an increasing
function of $V$ is the same pretty argument as before. 
  
\section{Remarks} \label{Conc}

We start by briefly mentioning other possible applications. The list
is clearly incomplete.

\begin{itemize}
\item $X$ can be a drug that interacts with specific cells, so this is
  a suitable framework for chemotherapy modelling;
\item $X$ may be bacteria that are ingested by neutrophiles, and can
  multiply inside the cell. This framework is therefore a possible
  model of bacterial infection and resistance \cite{MSR};
\item $X$ can be a mitogen, giving a model of stem cell number
  maintenance in which cells that ingest enough mitogen will multiply,
  and divide their mitogen label among the daughter cells, while cells
  that do not have enough mitogen, die~\cite{Kitadata}.
\end{itemize}

Clearly the $i$-equation for $y$ could have been an stochastic
differential equation; then the $x$-equation would have been an
integro-stochastic differential equation, and the $M$ equation a
stochastic partial differential equation. Furthermore, extensions to
multidimensional $x$ and $y$ are straightforward although
incorporating spatial structure seems to us much more challenging (as
it is for coagulation--fragmentation equations).

To summarise, we have presented a modelling framework that seems to
cover a vast number of possible modelling contexts beyond the reach of
coagulation--fragmentation equations. Such a framework necessitates
the analysis of complicated mathematical objects and so work on
existence, structure of equilibria, convergence to equilibria and
their regularity, etc., is required when specific examples are
considered. The search will be for measure-valued solutions, and
relevant work in this direction has been undertaken by Carrillo,
Gwiazda and co-workers; see, for example, \cite{CCGU, GM}. In terms of
possible numerical solutions to the governing equations, there are no
off-the-shelf numerical methods that we know of, although it seems
that escalator box train (EBT) methods could be adapted to the problem
(see related work of Carrillo, Gwiazda and Ulikowska \cite{CGU}).

Finally, we note that in many of the biological settings for which
this framework could be used, deeper understanding of the active
transport of molecules in and out of cells may be needed. Good models
of transport across membranes (facilitated transport, phagocytosis,
pumps etc.) are relatively sparse in the literature (though see, e.g.,
\cite{Naftalin,Rea}) and further work on such models would be of
significant benefit.

\end{document}